\documentclass[aps,prl,twocolumn,showpacs,letterpaper,superscriptaddress]{revtex4}
\usepackage{graphicx}
\usepackage{amsmath}

\begin{document}
\title{High-energy excitonic effects in graphite and graphene}

\author{Paolo E. \surname{Trevisanutto}}
\affiliation{Institut N\'eel, CNRS \& UJF, Grenoble, France}
\affiliation{D\'epartment de physique and RQMP, Universit\'e de Montr\'eal, Canada}
\affiliation{European Theoretical Spectroscopy Facility (ETSF), France}

\author{Markus \surname{Holzmann}}
\affiliation{LPMMC, CNRS \& UJF, Grenoble and LPTMC, CNRS \& UPMC, Paris, France}
\affiliation{European Theoretical Spectroscopy Facility (ETSF), France}

\author{Michel \surname{C\^{o}t\'e}}
\affiliation{D\'epartment de physique and RQMP, Universit\'e de Montr\'eal, Canada}

\author{Valerio \surname{Olevano}}
\affiliation{Institut N\'eel, CNRS \& UJF, Grenoble, France}
\affiliation{European Theoretical Spectroscopy Facility (ETSF), France}

\date{\today}

\begin{abstract}
We present \emph{ab initio} many-body calculations of the optical absorption in bulk graphite, graphene and bilayer of graphene.
Electron-hole interaction is included solving the Bethe-Salpeter equation on top of a GW quasiparticle electronic structure.
For all three systems, we observe strong excitonic effects at high energy, well beyond the continuum of $\pi \to \pi^*$ transitions.
In graphite, these affect the onset of $\sigma \to \sigma^*$ transitions.
In graphene, we predict an \textit{excitonic resonance} at 8.3 eV arising from a background continuum of dipole forbidden transitions.
In the graphene bilayer, the resonance is shifted to 9.6 eV.
Our results for graphite are in good agreement with experiments.
\end{abstract}

\maketitle
Graphene, a recently discovered 2D hexagonal crystal carbon sheet \cite{Novoselov}, has attracted much interest due to its exotic electronic properties.
The comparison with ordinary 3D graphite, the ABA stacking of graphene layers with weak inter-layer interactions,
gives further insights, as screening effects and effective 2D confinement are modified there.
Both systems share a peculiar semi-metallic character together with a strong electronic anisotropy,
giving rise to
optical properties of particular interest, especially in view of technological applications in opto-electronics \cite{optoelectronics}.
More generally, also in astrophysics, accurate determination of
optical absorption of carbon structures is  fundamental \cite{Papoular}.

The optical absorption of graphite was experimentally determined by reflectance \cite{TaftPhilipp,KluckerSkibowski} as well as by energy-loss via Kramers-Kr\" onig \cite{Zeppenfeld,TosattiBassani,Venghaus,Buchner}, showing important anisotropy effects between measurements using polarizations parallel or perpendicular to the crystallographic \textit{c} axis.
Optical properties of graphene are under current experimental investigation \cite{graphene-opt,Nair,Mak}.
A first theoretical JDOS (joint density of states) calculation \cite{PastoriParra}, based on an independent particle picture and usual selection rules, has shown, that the optical absorption spectrum of graphite can be divided in two regions: the visible range from  0 eV up to 5 eV originates from transitions among $\pi$ bands, whereas the region beyond 10 eV is made of $\sigma$ band transitions.
\textit{Ab initio} calculations beyond JDOS and in the RPA approximation have been done on graphite \cite{AhujaAuluck} and  graphene, also including local-field effects \cite{Marinopoulos}.

In this work, we extend previous theoretical results \cite{AhujaAuluck,Marinopoulos} beyond RPA by including electron-electron (\textit{e-e}) and electron-hole (\textit{e-h}) interactions.
Our calculations are based on the many-body \emph{ab initio} GW and Bethe-Salpeter equation (BSE) approach \cite{OnidaReining}.
We study bulk graphite, free-standing graphene, and a bilayer of graphene, considering both polarizations.
With respect to Ref.~\cite{Cohen&Louie} which is restricted to the low-energy range, our work extends also to high energies.

We observe strong excitonic effects at unusual high energies, well beyond the continuum of $\pi$-$\pi^{*}$ transitions.
In graphite, excitonic effects strengthen the low energy part of the $\sigma$-$\sigma^*$ structure and reshape it.
Both, in the perpendicular and in the parallel polarization, our spectra are in very good agreement with experiment, in particular with Ref.~\cite{Venghaus}.
In graphene, we  predict an intense peak at 8.3 eV that we interpret as an \textit{excitonic resonance} arising from a background single-particle continuum of dipole forbidden transitions.
In the bilayer the resonance remains, but it is shifted to 9.6 eV due to reduced confinement and increased screening.
Measuring optical spectra, this feature could thus be used as a fingerprint to discriminate between graphene and multi-layer graphene.

Our GW and BSE calculations  \cite{OnidaReining} are based on ground state calculations using density-functional theory in the local-density approximation (DFT-LDA).
Starting from the Kohn-Sham DFT electronic structure, we calculate an \textit{ab initio} GW quasiparticle electronic structure that takes into account \textit{e-e} many-body interactions.
In order to include \textit{e-h} interactions in the response functions, we solve the Bethe-Salpeter equation for the two-particle correlation function $L$,
\begin{equation}
L = GG + GG \Xi L
,
\label{BSE}
\end{equation}
where $G$ is the GW Green's function, and $\Xi$ is  the BSE kernel. We have used the
approximation $\Xi = -iv + iW$, where $v$ is the Coulomb and $W$ the screened interaction.
As described in Ref.~\cite{OnidaReining}, we define $H^\textrm{exc} = (\epsilon^\textrm{GW}_c - \epsilon^\textrm{GW}_v) + \Xi$, and remap the BSE, Eq.~(\ref{BSE}), into a 2-particle Schr\" odinger equation
\begin{equation}
H^\textrm{exc} \Psi^\textrm{exc}_\lambda = E^\textrm{exc}_\lambda \Psi^\textrm{exc}_\lambda
,
\label{BSE-2}
\end{equation}
where $E^\textrm{exc}_\lambda$ represent the excitation energies including the \textit{e-h} interaction effects, and  $\Psi^\textrm{exc}_\lambda$ the excitonic wavefunctions.
We diagonalize Eq.~(\ref{BSE-2}) working with a basis of Kohn-Sham bilinears, $\Psi^\textrm{exc}_\lambda(r_h,r_e) = \sum_{kvc} \Psi^{kvc}_\lambda \phi^\textrm{KS*}_{vk}(r_h) \phi^\textrm{KS}_{ck}(r_e)$ where
$v$ ($c$) runs on valence (conduction) bands, and $k$ lies in the 1st Brillouin zone.
The macroscopic dielectric function is then given by
\[
\varepsilon(\omega) = 1 -
  \lim_{q \to 0} v(q) \sum_\lambda \frac{\big\vert \sum_{kvc} \Psi_\lambda^{kvc}
           \langle \phi^\textrm{KS}_{vk+q}|e^{-iqr}| \phi^\textrm{KS}_{ck}\rangle \big\vert^2}
           {E^\textrm{exc}_\lambda-\omega +i\eta}.
\label{diel_tens}
\]

\begin{figure}
 \includegraphics[clip,width=0.45\textwidth]{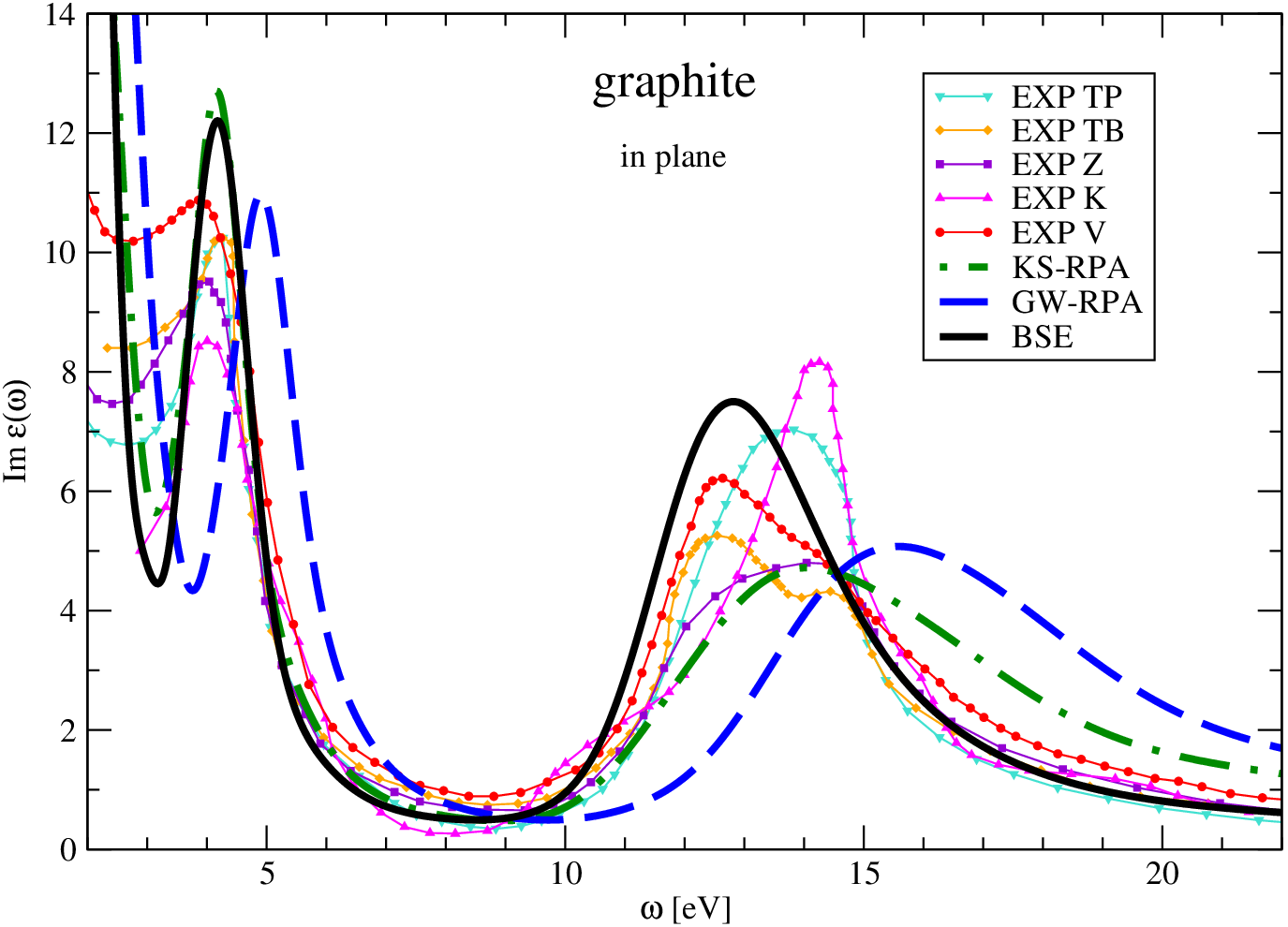}
 \caption{(color online) Optical absorption spectra of bulk graphite for $\mathbf{E}\perp\mathbf{c}$. Solid black line: BSE; blue dashed line: GW-RPA; green dot-dashed line: KS-RPA. All theoretical curves convoluted by a relative Gaussian broadening of $\sigma=0.075 \, \omega$. Experiments: TP, cyan triangles down \cite{TaftPhilipp}; TB, orange diamonds: \cite{TosattiBassani}; Z, indigo squares: \cite{Zeppenfeld}; K, magenta triangles up: \cite{KluckerSkibowski}; V, red circles: \cite{Venghaus}.}
 \label{graphitexy}
\end{figure}

The ground-state DFT-LDA and the GW corrections have been calculated using the \texttt{ABINIT} code.
In the case of the bilayer and of graphene, the layers are isolated by 38 Bohr of vacuum, distance large enough to avoid spurious interactions between replicas.
We used Martins-Trouiller pseudopotentials with \textit{s} and \textit{p} electrons in the valence.
The BSE calculation was carried out by the \texttt{EXC} code.
For graphite, the Brillouin zone was sampled with a (8 8 5)  Monkhorst-Pack k-point grid shifted by (0.01, 0.02, 0.03), whereas a (16 16 1) shifted grid was used for both graphene and bilayer.
Wavefunctions have been represented using 967 plane waves (39 Ry) for graphite and 1367 (24 Ry) for the bilayer and graphene, while the dimension of the kernel $\Xi$ was 53 (graphite) and 287 (bilayer and graphene) plane waves (8 Ry).
We included all $\sigma$ and $\pi$ occupied bands, i.e. 8 in graphite and the bilayer, 4 in graphene.
And 13 empty bands in graphite, 25 in the bilayer and 21 in graphene.
We also used the \texttt{DP} code to obtain random-phase approximation (RPA) spectra.
Local-field effects have been taken into account, both in RPA, and in BSE calculations.
All theoretical curves have been convoluted by a Lorentzian with a quadratic width $\eta=0.003 \, \omega^2$ (Fermi liquid-like lifetime behavior) to introduce the intrinsic quantum broadening, together with a Gaussian broadening adapted for the presumed experimental energy resolution.


\begin{figure}
 \includegraphics[clip,width=0.45\textwidth]{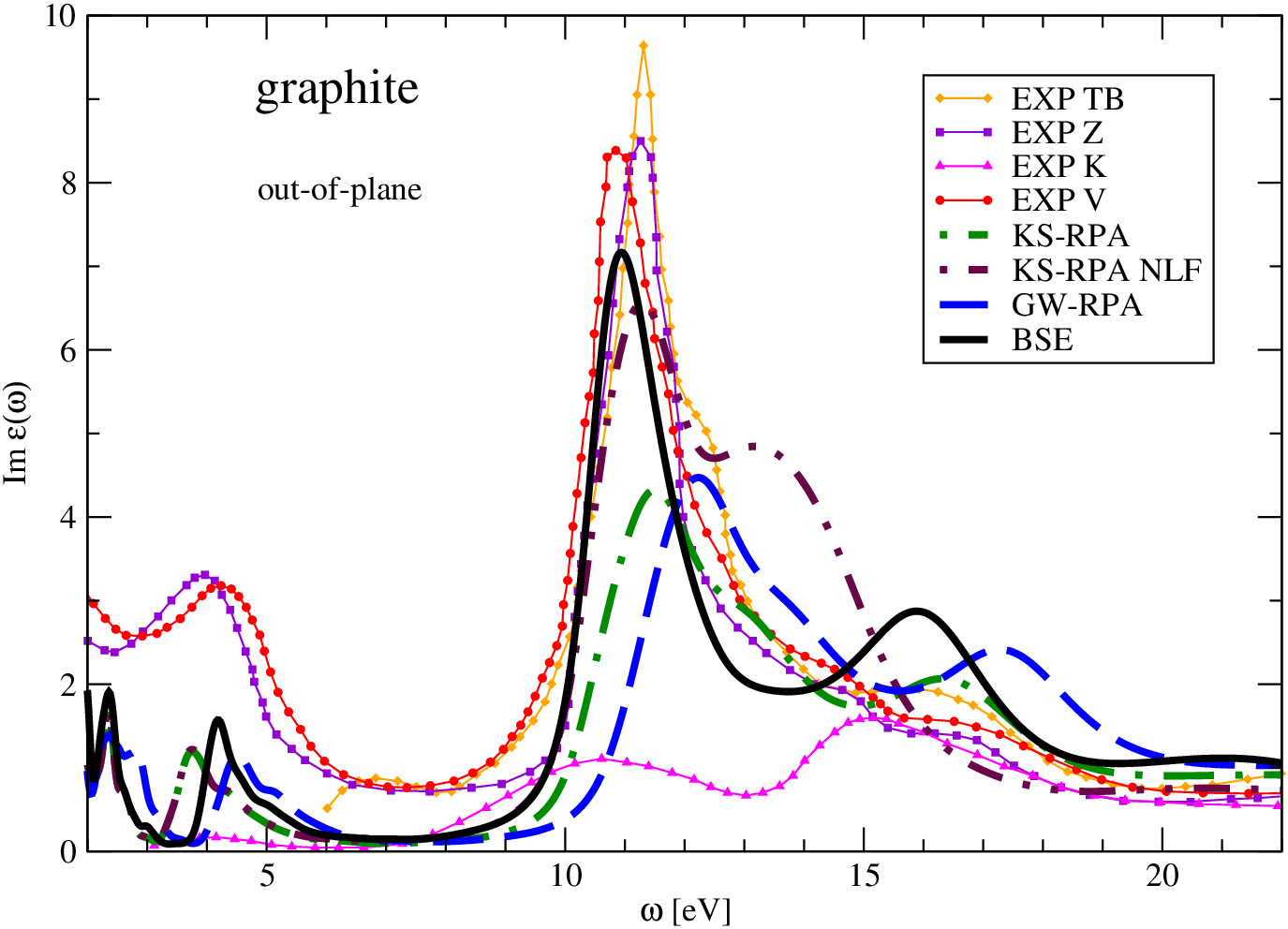}
 \caption{(color online) Optical absorption spectra of graphite for $\mathbf{E}\parallel\mathbf{c}$. Same notation as in Fig. \ref{graphitexy}, but with a relative Gaussian broadening of $\sigma=0.025 \, \omega$.}
\label{graphitez}
\end{figure}

\textit{Graphite, in-plane polarization:}
The calculated imaginary part of the macroscopic dielectric function, directly associated with the absorption spectra, is shown in Fig.~\ref{graphitexy} for graphite in the $\textbf{E}\perp\textbf{c}$ polarization.
The Kohn-Sham RPA (KS-RPA) spectra reproduces previous theoretical results \cite{Marinopoulos} calculated using a denser grid of k-points.
Local-field (LF) effects are expected to be small since the in-plane electronic density of graphene is almost homogeneous \cite{Marinopoulos}.
Indeed, spectra with and without LF coincide for this polarization.
The position of the lowest energy 4.2 eV peak of the KS-RPA spectra is exactly the same as in all the considered experiments.
On the other hand, the position of the highest energy structure due to $\sigma \rightarrow \sigma^{*}$ is compatible with only some of the experiments.

The inclusion of \textit{e-e} GW effects  globaly shifts the spectrum to higher energies.
The shift however is not rigid.
GW corrections are less effective on bands closer to the Fermi energy around K,
but they increase with energy, and correct the band-gap underestimation of the DFT-LDA electronic structure, for example at $\Gamma$ and L, resulting in an increased agreement with photoemission experiments. 
However, concerning optical absorption, the agreement with the experiment (see Fig.~\ref{graphitexy})
gets worse when passing from KS-RPA to GW.
For example, the KS-RPA main structure at 14 eV has been shifted to 15.3 eV by GW corrections, at least 1.3 eV off from the experimental position, and, similar, also  the lowest energy structure around 4 eV, shifted by 0.7 eV.

\begin{figure}
 \includegraphics[clip,width=0.45\textwidth]{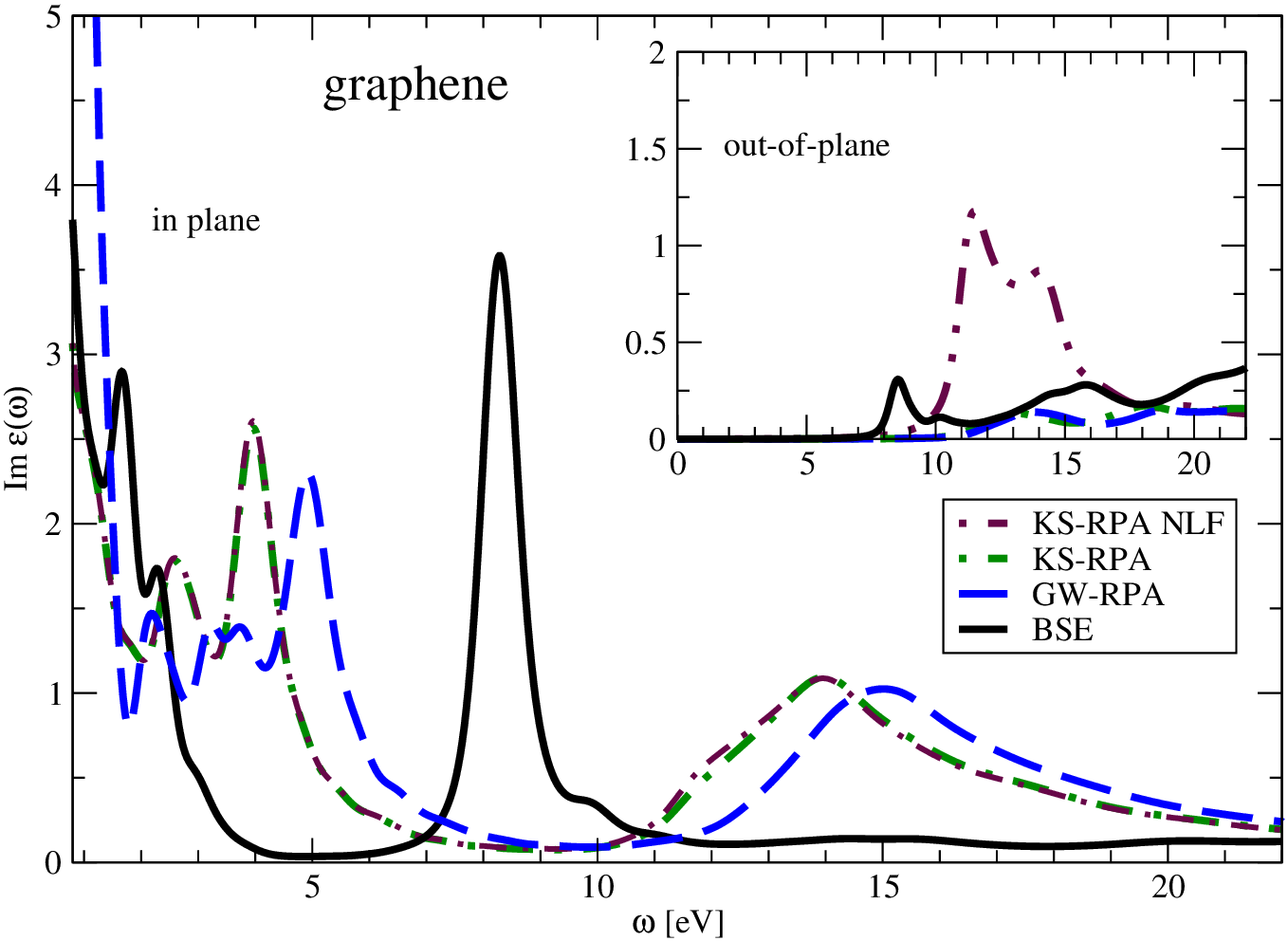}
 \caption{(color online) Optical absorption spectrum of graphene for $\mathbf{E}\perp\mathbf{c}$. Inset for $\mathbf{E}\parallel\mathbf{c}$. Same notation as in Fig. \ref{graphitexy} with an absolute Gaussian broadening of 0.2 eV.}
 \label{Graphene}
\end{figure}

The inclusion of \textit{e-h} interaction effects via BSE seems to compensate \textit{e-e} effects, restoring a good agreement with experiment, similar to the KS-RPA result.
The  low energy peak position agrees with all five considered experiments, although the height and the shape differ somehow.
The position of the high energy peak  recovers  a position at smaller energy than the KS-RPA peak.
Its profile is slightly reshaped, with a strengthening of the peak, particulary on the low-energy side.
With respect to KS or GW-RPA, the peak position appears now to be at 12.8 eV.
A fine analysis reveals a main excitonic energy at 12.6 eV, together with  two other main excitation energies at 13.3 and at 13.7 eV.
The latter should conjure an asymmetric, slower drop on the right shoulder of the peak.
Unfortunatly,  there is some disagreement between  experimental spectra in this high energy region measured by different techniques \cite{TaftPhilipp,KluckerSkibowski,Zeppenfeld,TosattiBassani,Venghaus}, mostly energy-loss and reflectivity measurements, and the position of the high-energy peak varies between 12.6 and 14.3 eV.
The position of the peak of our BSE calculation is more in agreement with the results of Venghaus \cite{Venghaus} and also Tosatti and Bassani \cite{TosattiBassani}, both at 12.6 eV and both measured by energy-loss, which do not suffer from surface effects that tamper the bulk result for reflection experiments.
The shoulder observed at $\sim 14$ eV  could correspond to  our excitation at 13.7 eV, although the imposed broadening (linear Gaussian with $\sigma=0.075 \, \omega$) makes it less evident in the figure.
The agreement of our result with the optical experiment of Taft and Philipp \cite{TaftPhilipp} is still acceptable, but worsens when comparing with Zeppenfeld \cite{Zeppenfeld} (peak at 14 eV) and in particular with Klucker et al. \cite{KluckerSkibowski} (peak at 14.3 eV), measured via X-ray absorption.
Comparing BSE with GW, it is remarkable that excitonic effects are stronger in the high-energy range of the spectrum rather than in the lower one,
in contrast to ordinary semiconductors or insulators where excitonic effects mainly affect the lowest energy part of optical spectra, or to metals, where excitonic effects are in general negligible.

\begin{figure}
 \includegraphics[clip,width=0.45\textwidth]{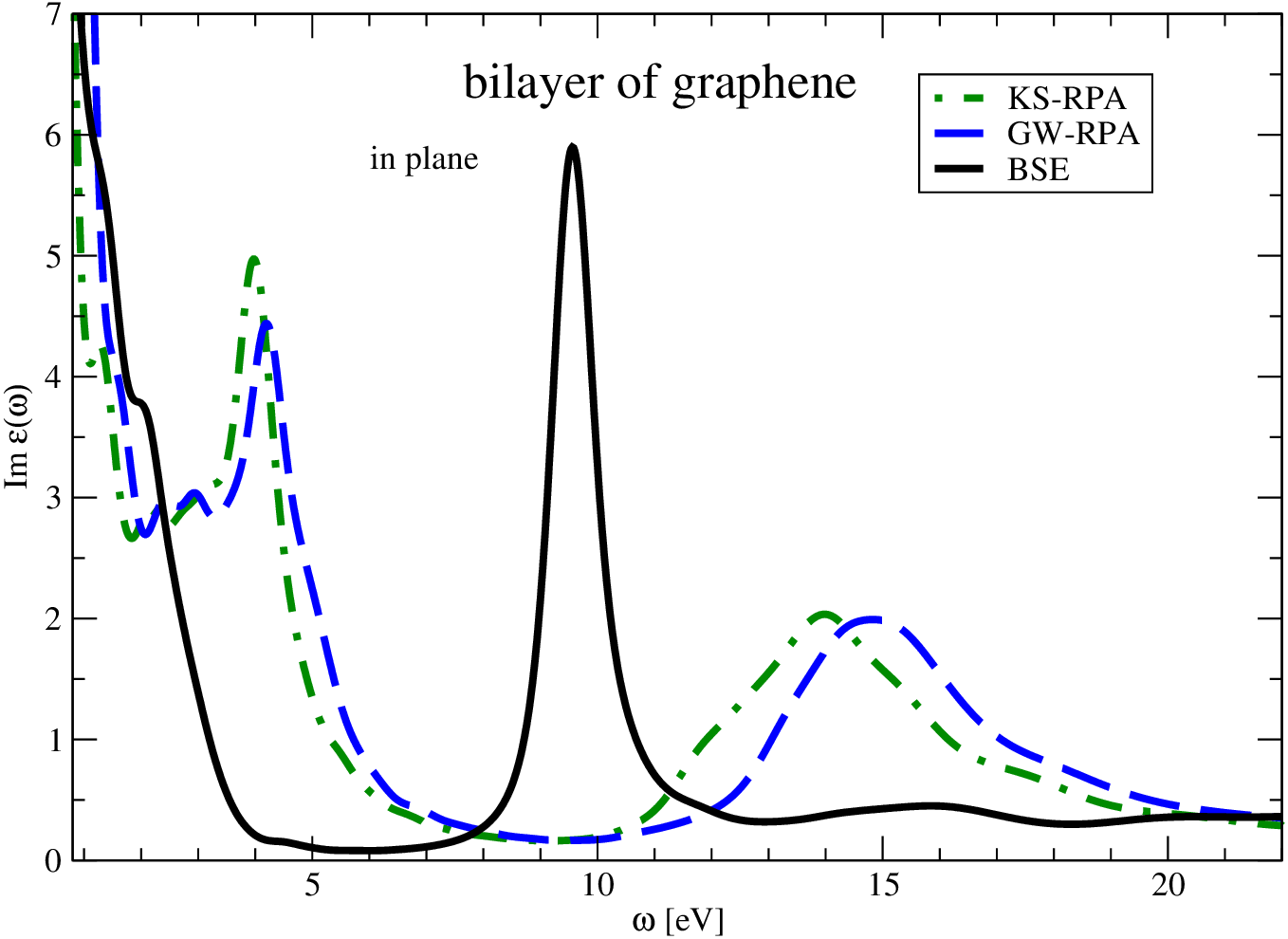}
 \caption{(color online) Optical absorption spectrum of bilayer of graphene for $\mathbf{E}\perp\mathbf{c}$. Same notation as in Fig. \ref{graphitexy} with an absolute Gaussian broadening of 0.2 eV.}
  \label{Bilayer}
\end{figure}

\textit{Graphite, out-of-plane polarization:}
Previous conclusions are further confirmed by the results for the $\textbf{E}\parallel\textbf{c}$ polarization (Fig.~\ref{graphitez}).
As already found in Ref.~\cite{Marinopoulos}, LF effects are stronger in the out-of-plane case, as it is evident  comparing  Kohn-Sham RPA  with and without LF.
Both curves are in agreement with previous theoretical results \cite{AhujaAuluck,Marinopoulos}.
The KS-RPA with LF curve is again in fair agreement with the experiment.
The position of the $3.7$ eV low energy peak is quite accurate, the main peak at $11.5$ eV not far from the experiment, the last peak at $16.3$ eV and a shoulder at $\sim 13.2$ eV in correspondence with experimental features.
Again, the inclusion of \textit{e-e} GW interaction effects causes a blue shift of up to 1 eV and worsens the agreement with the experiment.
The improvement appears only when including also \textit{e-h} interaction effects (BSE).
The peak at low energy is now between the KS-RPA and the GW-RPA peaks, at exactly 4.2 eV, in agreement with experiment, especially with Venghaus \cite{Venghaus}.
The main peak instead red-shifts beyond the KS-RPA by 0.6 eV and is placed at 10.9 eV, again in good agreement with Venghaus (10.8 eV) and at lower energy with respect to Zeppenfeld \cite{Zeppenfeld} and Tosatti and Bassani \cite{TosattiBassani} (both at 11.3 eV).
With respect to KS-RPA and GW, \textit{e-h} effects also enhance the peak, and improve the agreement with the experiment on the magnitude.
The higher energy peak gets  restored at the KS-RPA position with a slight improvement on the position, but a worsening in intensity.
The $\sim 13$ eV shoulder, still evident in KS-RPA, has been washed out by the strengthening of the neighbor main peak and is no longer visible.
Our actual BSE calculation supports the scenario supposed for many-body effects by Ref.~\cite{Marinopoulos}.

With respect to experiment, in the case of  out-of-plane polarization, the situation is similar  to the in-plane polarization.
The BSE result is in particular in good agreement with Venghaus \cite{Venghaus}, but also with Tosatti and Bassani \cite{TosattiBassani} and Zeppenfeld \cite{Zeppenfeld}, but disagrees  with Klucker et al. \cite{KluckerSkibowski}.
However,  apart from the peak positions, the agreement with the experiment can be improved concerning peak intensities.
In particular the low-energy part of the spectrum, due to interlayer interactions, still seems to be underestimated.
There is also a slight underestimation on the main peak, although for this polarization we used less broadening ($\sigma=0.025 \, \omega$) than in the in-plane polarization.

\textit{Graphene:}
In graphene (Fig.~\ref{Graphene}), due to its effective 2D dimensionality,  LF effects become the most important effects for the out-of-plane polarization.
As already found in Ref.~\cite{Marinopoulos}, for $\mathbf{E} \parallel \mathbf{c}$, LF effects produce a complete suppression of spectra due to a depolarization effect (compare KS-RPA with and without LF curves in the inset of Fig.~\ref{Graphene}). The 2D selection rules \cite{PastoriParra} forbid  $\pi \to \pi^*$ transitions ($<$10 eV), allowing  $\sigma \to \sigma^*$ ones instead. Nevertheless, the strong anisotropy increases the LF effects suppressing the latter transitions as well.
On the other hand, LF effects are negligible in-plane (Fig.~\ref{Graphene}).
In graphene as in graphite, GW \textit{e-e} interaction effects produce a $\sim 1$ eV blue shift of the KS spectrum.
Both the $\pi \to \pi^*$ peak at 4 eV and the $\sigma \to \sigma^*$ peak at 14 eV are shifted around $+1$ eV.
In graphene, however, the inclusion of \textit{e-h} interaction turns out to cause a much more spectacular effect.
The spectrum is completely reshaped.
Surprisingly, at $4 \sim 5$ eV, as well as at $14 \sim 15$ eV, there is not anymore any oscillator strength.
The RPA peaks disappear completely.
Most of the spectral weight is pushed to very low energies, $< 3$ eV.
However, unexpectedly, a strong peak rises up at 8.3 eV in the region where single-particle oscillator strengths were vanishing.
\emph{This excitonic resonance arises from a background single-particle continuum of dipole forbidden transitions.}
An analysis over the excitonic oscillator strengths $|\Psi^{vck}_\lambda \langle vk+q|e^{-iqr}|ck\rangle|$ relative to the most intense excitonic eigenvalue indicates that the strong excitonic resonance is conjured by an extended mixing of  $\sigma\rightarrow \sigma^{*}$ transitions over a large energy range, from 12.5 to 17 eV.
An excitonic effect at so large distance in energy (8.3 vs 12.5-17 eV) is caused by the increased \textit{e-h} interaction due to reduced screening in graphene.
Since we do not have any absorption spectrum experiments in graphene for comparison, we maintained the broadening at the minimum level to highlight these effects.
Depending on the experimental energy resolution, this peak could appear more broadened, but does not affect
the peak  position at {$< 8.5$ eV, and a  slight right asymmetry.
We note that inclusion of band 2 in the BSE is the only crucial ingredient to start to conjure the high energy excitonic resonance, whereas the $4 \sim 5$ eV \cite{Cohen&Louie} energy region is very sensitive to the total number of bands, the dimension of the BSE kernel, and the vacuum distance between graphene layers.


\textit{Bilayer:}
Similar to graphene, an excitonic resonance also occurs for the graphene bilayer (Fig.~\ref{Bilayer}), however, at larger energy, 9.6 eV. 
The difference is due to the increased screening with respect to graphene, which reduces \textit{e-h} interaction strength.
With respect to graphene, RPA spectra appear different at low energy due to interlayer effects.
However, also here, \textit{e-h} interaction effects sweep the region at $\sim$ 5 eV, and push the oscillator strength towards lower energies $< 3$ eV, or towards the excitonic resonance.

In conclusions, we have presented \emph{ab initio} GW-BSE calculations of the optical absorption spectra in graphite, graphene, and the bilayer, taking into account \textit{e-e}, \textit{e-h}, and LF effects.
For the three systems, we have found strong excitonic effects in a high-energy range $> 8$ eV, well beyond the range of $\pi \to \pi^{*}$ transitions.
In graphene and bilayer, these give rise to an excitonic resonance on a background continuum of dipole forbidden transitions.
The excitonic resonance has a main $\sigma \to \sigma^*$ character.
Our BSE spectra on graphite are in good agreement with experiment.


We thank L. Reining, G. Martinez and M. Ladisa for useful discussions.
P.E.T. was supported by ANR ETSF France.
Computer time was granted by Ciment and we acknowledge the \textit{Fondation Nanoscience} for support via the RTRA NanoSTAR, Dispograph and Muscade projects.
We have used \texttt{EXC}, \texttt{ABINIT} and \texttt{DP} codes.


\begin{thebibliography}{99}

\bibitem{Novoselov}
S.~Novoselov et al. Nature {\bf 438}, 197 (2005).

\bibitem{optoelectronics}
Ph.~Avouris, Z.~Chen, and V. Perebeinos, Nature Nanotechnology \textbf{2}, 605 (2007).

\bibitem{Papoular}
R. J. Papoular and R. Papoular, Mon. Not. R. Astron. Soc. \textbf{394}, 2175 (2009).




\bibitem{TaftPhilipp}
E. A. Taft and H. R. Philipp, Phys. Rev. \textbf{138}, A197 (1965).

\bibitem{KluckerSkibowski}
R. Klucker, M. Skibowski and W. Steinmann, Phys. Stat. Sol. (B) \textbf{66}, 703 (1974).

\bibitem{Zeppenfeld}
K. Zeppenfeld, Z. Phys. \textbf{211}, 391 (1968).

\bibitem{TosattiBassani}
E. Tosatti and F. Bassani, Nuovo Cimento B \textbf{65}, 161 (1970).

\bibitem{Venghaus}
H. Venghaus, Phys. Stat. Sol. (B) \textbf{71}, 609 (1975).

\bibitem{Buchner}
U. B\"{u}chner, Phys. Stat. Sol. (B) \textbf{81}, 227 (1977).

\bibitem{graphene-opt}
S. Roddaro \textit{et al.}, NanoLetters, \textbf{7} 2707 (2007).

\bibitem{Nair}
R. R. Nair \textit{et al.}, Science \textbf{320}, 1308 (2008).

\bibitem{Mak}
K. F. Mak \textit{et al.},\textbf{101}, 196405 (2008).

\bibitem{PastoriParra}
F. Bassani and G. Pastori Parravicini, Nuovo Cimento B \textbf{50}, 95 (1967).

\bibitem{AhujaAuluck}
R. Ahuja \textit{et al.}, Phys. Rev. B \textbf{55}, 4999 (1997).

\bibitem{Marinopoulos}
A.~G.~Marinopoulos, L. Reining, A. Rubio and V. Olevano, Phys. Rev. B {\bf 69}, 245419 (2004).

\bibitem{Cohen&Louie}
L. Yang \textit{et al.}, Phys. Rev. Lett \textbf{103}, 186802 (2009).

\bibitem{OnidaReining}
G. Onida, L. Reining and A. Rubio, Rev. Mod. Phys. \textbf{74}, 601 (2002) and references therein.











\end{thebibliography}
\end{document}